# Control of ribosome traffic by position-dependent choice of synonymous codons

**Namiko Mitarai*[1] and Steen Pedersen[2]**
[1] Niels Bohr Institute, University of Copenhagen, Blegdamsvej 17, DK-2100 Copenhagen Ø, Denmark.
[2] Department of Biology, University of Copenhagen, Ole Maaløes vej 5, DK-2200 Copenhagen N, Denmark.
* Corresponding Author ; E-mail: mitarai@nbi.dk

**Abstract.** Messenger RNA encodes a sequence of amino acids by using codons. For most amino acids there are multiple synonymous codons that can encode the amino acid. The translation speed can vary from one codon to another, thus there is room for changing the ribosome speed while keeping the amino acid sequence and hence the resulting protein. Recently, it has been noticed that the choice of the synonymous codon, via the resulting distribution of slow- and fast-translated codons, affects not only on the average speed of one ribosome translating the messenger RNA (mRNA) but also might have an effect on nearby ribosomes by affecting the appearance of "traffic jams" where multiple ribosomes collide and form queues. To test this "context effect" further, we here investigate the effect of the sequence of synonymous codons on the ribosome traffic by using a ribosome traffic model with codon-dependent rates, estimated from experiments. We compare the ribosome traffic on wild type sequences and sequences where the synonymous codons were swapped randomly. By simulating translation of 87 genes, we demonstrate that the wild type sequences, especially those with a high bias in codon usage, tend to have the ability to reduce ribosome collisions, hence optimizing the cellular investment in the translation apparatus. The magnitude of such reduction of the translation time might have a significant impact on the cellular growth rate and thereby have importance for the survival of the species.

## 1. Introduction

Messenger RNA (mRNA) encodes a sequence of amino acids by using triplets of nucleotides (codons). Except for two cases, methionin and tryptophan, more than one codon may encode the amino acid (synonymous codons). It is well known that one or more of the synonymous codons often are preferred over the use of the other codons for the same amino acid and that such bias depends on the gene, and that the highly expressed genes have a stronger bias in codon usage [1, 2, 3, 4, 5].

The codon bias indicates that the synonymous codons are not equal in function. Because stronger bias is observed in more highly expressed genes [5,6,7], which account for a large fraction of total protein and needs to be expressed under many growth conditions, it has been conjectured that the codons used in these genes are translated more efficiently. In order to address the relation between the codon usage and the protein expression, the Codon Adaptation Index (CAI) [5] was proposed. CAI characterizes the extent of codon usage bias of a gene in comparison to the codon usage in very highly expressed genes. A correlation between high CAI and a fast translation rate of the mRNA was established [8, 9]. Note that the faster translation speed does not mean higher production of protein, because the initiation rate of the ribosomes often is the rate-limiting step. The selection pressure is more likely due to an increase in the global ribosomal efficiency by increasing the turnover rate of ribosomes [2, 6, 7].

The codon dependent translation rate has been measured for four individual codons [10]. Based on these and a number of translation time determinations for individual mRNA or inserted sequences [8] a preliminary table for the codon specific translation rates were proposed [11] where the codons that are used more often in high CAI genes are translated faster than average. There are probably several physical causes of the difference in rates, for example the difference in the transfer RNA (tRNA) concentration, the degree of tRNA charging and/or the strength of the interaction between tRNA and codon in the ribosomal A-site [10,12].

Here, we study the effect of the position-dependent choice of synonymous codon on the ribosome traffic in *Escherichia coli*, by using the model of ribosome traffic with codon-dependent rates based on the proposed translation rates [11]. We especially focus on the ability of the codon choice to reduce the translation time by regulating ribosome collisions and queues. It is observed that some mRNAs have a less biased codon usage at the beginning of the mRNA compared to the distal part [7,13]. We previously proposed that such less biased sequence can reduce ribosome collisions and the following "traffic jams" by occluding the start site for instance when ribosomes for stochastic reasons initiate too often (Fig. 1). This increases the ribosomal efficiency significantly, because collisions slow down the ribosomes and might trap many ribosomes in a queue for a long time [11]. Similar scenario has been proposed later for yeast and other higher organisms [14, 15]. We furthermore argued that the less biased codon usage in the beginning of genes was selected for in the evolution rather than that these had drifted towards less biased codons due to an absence of a selection pressure for specific codons in this region [11]. Note that regulating traffic is different from just having a higher average speed: Even if most of the codons are translated fast, if there is a slow codon in the later part of the gene, queuing can evolve if the ribosomes initiate translation often enough.

In this paper, we quantify the ability of the codon usage to regulate ribosome traffic for several natural genes. We first demonstrate the significance of the location of a slow codon on the translation time by simulating sequences where the location of the slow codon is systematically changed. We then compare translation of the wild type (WT) sequence with sequences where the positions of the synonymous codons are randomly swapped within the gene. We show that the genes with high CAI tend to be better in preventing ribosome collisions and quantify how favorable such reduction of the translation time in the WT sequences is compared to the randomized sequences.

## 2. Model

First we describe the ribosome traffic model [11], which is based on simple lattice models of traffic flow [16, 17, 18, 19]. Figure 2 depicts ribosomes that translate a mRNA. In the model, a mRNA is modeled as one dimensional lattice, and one site corresponds to one codon. The system size $L$ corresponds to the total number of codons of the modeled gene. Translation starts by binding a ribosome 30S subunit to codon 1. Ribosomes try to bind with on-rate $K_s$, but the binding can occur only if the binding site is accessible, which requires that no ribosome is within the occluding distance $d = 11$ codons from the binding site. The ribosome stays at codon 1 for a waiting time, $\tau \sim 0.2$ s, to assemble the translating 70S ribosome. Then, all ribosomes try to move stochastically to a codon forward at a translation rate $R_x$ for the codon at the position $x$, where the ribosome is translating. Possible movements are accepted only if the distance to the preceding ribosome is larger than $d$. When the ribosome reaches the stop codon, the peptide is released with a rate $K_t$ and the ribosome leaves the mRNA. Parallel update with time step $dt = 1/300$ sec was used.

In this paper, we do not consider the mRNA degradation, since the mRNA lifetime is not always known. All the presented data are from the ribosomes translating in steady state.

We performed stochastic simulations of this model for several mRNA. The sequences studied in this paper are all *E. coli K12* genes and obtained from the database RegulonDB [20].

The translation rate $R_x$ at a position $x$ depends on the codon at this position. We use the codon-dependent rates determined in ref. [11], as summarized in table 1. To simplify the model, we assign one of the following three rates to each codon; A (~35 codons/sec), B (~8 codons/sec), and C-rate (~4.5 codons/sec). For most of the amino acids, the codon that is predominantly used in the genes encoding ribosomal proteins is assigned the fast A-rate except for proline: the rate for its most dominant codon CCG was measured to be moderate speed (~7 codons/sec), thus CCG was assigned middle B-rate and the rest of proline codons are assigned to be slow C-rate. Detailed description of the determination of the rate is given in ref. [11]. The termination rate $K_t$ is assumed to be 2/sec [11] and not rate-limiting.

We study various values of the on-rate $K_s$. The on-rate $K_s$ for *lacZ* for cells growing in a minimal media has been estimated to be about 0.9/sec [11]. Note that $K_s$ is larger than the actual initiation rate, since the occlusion time of the start codon is finite. In case of *lacZ*, this is about 1 sec, hence the actual initiation rate is about 0.5 /sec. In the following, we vary $K_s$ around this value from 1/2 to 2 fold. Note that the on-rate is subject to change depending on the condition of the cell, for example the quality of the growth media and the concentration of ribosomes. $K_s$ also depends on the ribosome binding site as well as the start codon, which varies depending on the gene [21]. $K_s$ is an important parameter for the ribosome traffic; when $K_s$ is low there will be few collisions between ribosomes, while it is likely that a high $K_s$ will cause collisions and ribosome queuing.

In order to see how good WT sequences are at regulating traffic jam, we simulate the following three cases for each gene:
- "WT": The sequence in the wild type gene.
- "Swap": The sequence that gives the same amino acid sequence, but the synonymous codons are randomly shuffled within the gene. This enables us to keep not only the amino acid but also the set of codons used (hence also CAI), and we can still see the effect of the location of the slow and fast codons. We take 100 randomized sequences per gene.
- "Fast": The sequence where the codons are replaced with the fastest synonymous codon. This is a measure of how fast the translation can be for a given mRNA, if the codon usage bias were the strongest.

The difference of the average translation speed between "WT" and "Fast" will be bigger for the genes with low CAI. The comparison between the ensemble of "Swap" and "WT" gives us an idea about how the "WT" has evolved by spatially organizing the codon sequence to increase the translation speed for the given CAI.

## 3. Results
3.1. Difference between genes: examples

In order to demonstrate the difference between individual genes, we present a few examples in this section.

Figure 3 shows the analysis of artificial sequences. When all the codons have the A-rate (Fig. 3a),

the mRNA's coverage with ribosomes is rather low. Some structures seen around the start- and stop-codon reflect the ribosome-diameter (11 codons). When one C-rate codon is added at the beginning (Fig. 3b), it increases the coverage of the start codon to somewhat. This slightly reduces the initiation rate resulting in lower coverage and fewer collisions between ribosomes. As the C-rate codon is moved to the middle (Fig. 3c) and to the end (Fig. 3d), some increase of the coverage around and upstream for the C-codon is seen due to the formation of ribosome queues. Figure 3(e) quantifies the C-rate codon position-dependence of the translation time per codon (average time for a ribosome to translate the whole gene divided by the number of codons in the gene) with various values of $K_s$. As the position of the C-rate-codon is moved from the beginning to the later part of the sequence, the translation time grows because of the induced collisions and small queues around the C-rate codon.

The translation time with the C-rate codon close to the stop codon is higher than the translation time with the C-rate codon close to the start codon by about 50% for the high on-rate $K_s = 1.8/s$. Having the slow codon at the beginning, on the other hand, reduces the translation time slightly compared to the all A-rate codon sequence.

This will give a selection pressure against mutations to C-rate codons in the distal part of the gene, because such mutations will reduce the ribosomal efficiency significantly. The mutation to have a C-rate codon at the beginning, on the other hand, will not increase the translation time and sometimes even reduces it by lowering the collisions between ribosomes. Therefore, fewer slow codons are expected at the later part of genes and this selection pressure should be stronger for the highly expressed genes.

Note that the ribosome flux (the number of translations per second) in Fig. 3(f) increases smoothly with $K_s$ and shows smaller variation with the position of the C-rate codon. It shows a small occlusion effect when the position C-rate codon is within 11 codons (ribosome diameter), but when it is close to the stop codon the dependence is negligible. The ribosome flux is often studied to investigate the ribosome traffic [22], but the on-rate is often the primary determinant of the ribosome flux in the biologically plausible parameter range. The translation time is sensitive to the position of the slow codons and contributes to the ribosomal efficiency [5, 7, 11], which should give bigger selection pressure to the position of the slow codon. We therefore focus on the translation time in the following.

Next, we show the results for a few natural sequences as examples. The first example is the *lacZ* gene, in which we have studied the ribosome traffic on WT sequence in the previous paper [11] and concluded that the WT has the ability to reduce ribosome queuing by having slow codons at the beginning. The translation time per codon for *lacZ* gene (Fig.4) shows that "Fast" is translated much faster, reflecting the not so high CAI of the *lacZ* gene (~0.35). Between "WT" and "Swap", we see that the time to translate the *lacZ* mRNA starts to be significantly different already at the biologically relevant middle $K_s$ and at high $K_s$ "WT" is translated 12 % faster than "Swap". Figure 5(a) compare the translation time distribution for the ensemble "Swap" of randomized *lacZ* sequences with the moderate on-rate $K_s = 0.9/s$ (dashed line) and high on-rate $K_s$ of $1.8/s$ (solid lines). The translation times for "WT" and "Fast" sequences with the corresponding on-rates are shown by arrows. The distribution for both of these are narrow and the "WT" sequence is translated around average for $K_s = 0.9/s$. When $K_s = 1.8/s$, the distribution of the "Swap" translations shows a tail towards a long translation time, whereas the "WT" stays with a close to unaltered rate. The long-tail for the high on-rate is caused by the formation of ribosome queues: If the slow codons gets located in the later part by the randomization, this sequence gets a very long translation time. Note that the CAI is identical for all the "Swap" sequences for each gene. Our results therefore indicate that the ability of the "WT" *lacZ* mRNA to regulate the ribosome queue formation matters especially at high on-rates.

In Fig. 5(b), the analogous plot for *lac* repressor encoded by *lacI* is shown. The ability of *lacI* "WT" to reduce "traffic jams" is not significantly improved but just at average for both low and high on-rate $K_s$. The *lacI* gene has a low CAI ($\approx 0.29$), thus "Fast" is much better than the "WT" and "Swap" samples.

The third example, *rpsA*, encodes protein S1 in the 30S ribosomal subunit and has high a CAI ($\approx 0.78$). The "WT" sequence belongs to the faster end of the distribution of "Swap" sequences for both values of the on-rate $K_s$. For high $K_s$, the "Swap" sequences show long tail to the slow translation time due to ribosome queuing, but the "WT" sequence stays at the fastest end of the distribution. The "Fast" sequence is still faster, but the difference is small compared to that of *lacZ* and *lacI* as expected from the higher CAI for *rpsA*.

3.2. Correlation between codon bias and the ability to reduce formation of ribosome queues.

The three examples in subsection 3.1 indicate that the genes with higher CAI tend to have better ability to reduce collisions and "traffic jams". The genes with higher CAI are often highly expressed, thus the reduction of the translation time in these genes is expected to be more important.

To investigate if this tendency is general, we studied an additional 84 genes. We chose genes with various CAI based on the categories in ref. [23]. The categories are: (i) Very highly expressed genes (27 genes) (ii) Highly expressed genes (13 genes) (iii) genes with moderate codon usage bias (20 genes), (iv) genes with low codon usage bias (19 genes), (v) regulatory / Repressor genes (8 genes). Table 2 lists these 87 genes studied here with the CAI value for each gene.
Figure 6 shows the summary of all these results. We plot as ordinate the translational efficiency of the wild type sequence, defined as the ratio $R$=(translation time for the "WT")/(Median of the translation times for the 100 randomized samples in "Swap") both determined for the biologically most plausible on-rate, $K_s = 0.9$/s. The abscissa is the CAI. If the translational efficiency ratio $R$ is less than one, it means that wild type sequence translated faster than the median of the randomized sequences and *vise versa*.

We use the median translation rate and not the average in the calculation of the translational efficiency ratio $R$, because of the long tail of the distribution when ribosome queues can be formed (cf. Fig. 5b); the average can be much bigger than the median. Using the average would then bias the result towards the conclusion that the "WT" sequence is better.

We see that there is a clear tendency for $R$ to become less than one for high CAI genes. This is perhaps more clearly depicted in the figure inset where the average $R$ for each group is shown. For $0.2<$CAI$<0.4$ (30 genes in total) 21 genes has $R< 1$, while for $0.4<$CAI$<0.6$ (28 genes in total) 5 genes has $R < 1$, and for $0.6 <$CAI (29 genes in total) 5 genes has $R < 1$. This indicates that there is a bias to have $R < 1$ (i.e. WT is better than the median) for CAI$> 0.4$ (Binomial test, $p < 0.01$). The reduction of the translation time for high enough CAI genes is about 2% on average, and some genes can get an about 5% reduction.

The effect strongly depends on the on-rate $K_s$. When the on-rate $K_s$ is doubled, the contrast between genes with high and low CAI becomes stronger and some genes can get a 15% reduction of the translation time compared to the median (data not shown). The on-rate might be smaller than the value for *lacZ* [11,24] especially for the genes with low expression (e.g. [25]). In such a case, the collisions between ribosomes become less frequent and the effect of the position of the slow codons will be weakened.

## 4. Summary and Discussion

We investigated the effect of changing the sequences of synonymous codons on the ribosome traffic by numerical simulations. After demonstrating that the position of one slow codon can give sizeable variation of the translation time, we compared the wild type sequence (WT), sequences with position of synonymous codons randomized within the gene (Swap), and the sequence with the fastest synonymous codons (Fast). We confirmed that the ability to regulate the traffic jams varies among the genes. The analysis of 87 genes showed that the WT genes with CAI above 0.4 have higher ability to reduce ribosome queuing compared to the randomized sequences. We find that an approximately 2 % reduction in the translation time can be achieved by this ability to minimize queue formation even with the moderate on-rate.

This indicates that the selection pressure to have shorter translation time is stronger for the high CAI genes. This is complementary to the view that higher codon bias evolved to reduce the translation time in highly expressed genes to increase the ribosomal efficiency [5,7]: If the mutation to the slower codon happens at the position that increases the collisions and "traffic jams", it will be strongly selected against compared to mutations at earlier positions, resulting in sequences with better at regulating ribosome traffic for high CAI genes.

Assuming that the investment in translational machinery is important for determining the cellular growth rate, we may estimate the effect of a mutation that changes the translation time on the growth [6, 7]. If a mutation occurred in gene encoding a protein with a fraction $y$ of total protein, and if the mutation leads to fraction $x$ reduction of the translation time, the doubling time will be reduced by $xy$. For example, if $x = 2\%$ (average reduction with $K_s = 0.9$/s), $y = 0.5\%$, and the original doubling time were 1 hour, then doubling time will be 0.01% less for the mutant and it would take about 3.8 years for the mutant population to become 10-fold above the non-mutant population. This may not be a long time compared to the evolutionary time scale, and the advantage will increase sizably if the on-rate for the gene was larger.

We are aware that a few very highly expressed genes exhibits the translational efficiency ratio $R$ larger than 1 (Fig. 6). There can be a few possible explanations of this observation.

The first possibility is that this is an artifact of having simplified rates, where we categorize all the codons into ABC-rates, which has rather strong contrast to each other. Since highly expressed genes tend to have many A-rates, having a few B or C-rate codons in the later part of the gene can cause a significant jamming, while for low CAI case, if slow codons are evenly distributed the contrast in the rates are averaged out. Thus, we might see stronger effect of "traffic jams" for high CAI genes in the present simulation compared to a situation where could be some difference within the A-rate codons rates. This should be tested after more precise values for the rates become available.

In addition, regulation of "traffic jams" is not the only selection pressure for the location of slow codons. For example, the functional half-life of *lacZ* mRNA is known to depend on the codon usage in the early coding region, and this might be related to the coverage of mRNA by ribosome in an early region [26]. The length of the mRNA half-life should be subject to strong selection pressure, since it proportionally affects the protein expression.

For highly expressed genes, we previously found that the codons used in the 3' part were less conserved than the codons used in the 5' coding part of the mRNA. We argue here that having a non-biased codon usage in the 3' coding part of the mRNA is selected against, because slowly translated codons here would lead to ribosome collisions and queues that would reduce the translational capacity. The likely solution to this dilemma is of course that the use of slowly translated codons in the beginning are even more selected for than the selection for the fast codons in the 3'coding part of the mRNA.

The codon usage in the beginning of the mRNA should 1) give a suitable mRNA half-life 2) reduce queuing in the later part of the mRNA. 2) may be supported further by the fine-tuning of the initiation rate via occlusion time due to slow codons [11]: If the translation rates for rare codons are unsaturated, the occlusion time can respond to the changes in the substrate level. Furthermore, secondary mRNA structures near the ribosome binding site is avoided [27] probably because such would reduce the ribosome on-rate, $K_s$ [28, 29, 30]. It is not obvious that slowly translated codons are less likely to form structured mRNA, but altogether it is clear that numerous regards should be solved by the codon usage in the beginning of the mRNA.

**Acknowledgments**
The authors thank Kim Sneppen for valuable discussions. This work was funded by the Danish National Research Foundation through Center for Models of Life.

**Table 1.** Codon and assigned translation rates. The column "aa" shows the corresponding amino acid for the given codon in the column "Codon", The column "Rate" gives the rate we assign based on ref.[10], A-rate: 35 codons/sec, B-rate: 8 codons/sec, and C-rate: 4.5 codons/sec. For the stop codons the termination rate $K_t$ is given. We assume $K_t$ =2/sec.

| Codon | aa | Rate | Codon | aa | Rate | Codon | Aa | Rate | Codon | aa | Rate |
|---|---|---|---|---|---|---|---|---|---|---|---|
| UUU | Phe | B | UCU | Ser | A | UAU | Tyr | B | UGU | Cys | B |
| UUC | Phe | A | UCC | Ser | A | UAC | Tyr | A | UGC | Cys | A |
| UUA | Leu | B | UCA | Ser | B | UAA | Stop | $K_t$ | UGA | Stop | $K_t$ |
| UUG | Leu | B | UCG | Ser | B | UAG | Stop | $K_t$ | UGG | Trp | A |
| CUU | Leu | B | CCU | Pro | C | CAU | His | B | CGU | Arg | A |
| CUC | Leu | B | CCC | Pro | C | CAC | His | A | CGC | Arg | A |
| CUA | Leu | B | CCA | Pro | C | CAA | Gln | B | CGA | Arg | C |
| CUG | Leu | A | CCG | Pro | B | CAG | Gln | A | CGG | Arg | C |
| AUU | Ile | A | ACU | Thr | A | AAU | Asn | B | AGU | Ser | B |
| AUC | Ile | A | ACC | Thr | A | AAC | Asn | A | AGC | Ser | B |
| AUA | Ile | B | ACA | Thr | B | AAA | Lys | A | AGA | Arg | C |
| AUG | Met | A | ACG | Thr | B | AAG | Lys | B | AGG | Arg | C |
| GUU | Val | A | GCU | Ala | A | GAU | Asp | B | GGU | Gly | A |
| GUC | Val | B | GCC | Ala | B | GAC | Asp | A | GGC | Gly | A |
| GUA | Val | A | GCA | Ala | A | GAA | Glu | A | GGA | Gly | C |
| GUG | Val | A | GCG | Ala | B | GAG | Glu | B | GGG | Gly | B |

**Table 2.** Names of the genes studied, with the CAI value for each gene. CAI is defined by using the Relative Synonymous Codon Usage (RSCU) for each codon characterizing the bias of the usage of the codon in highly expressed genes; CAI is the ratio between the geometric mean of RSCU's (GMR) of the gene and maximum possible value of GMR for the sequence of the amino acids of the given protein [5].

| Category | Gene names (CAI value) |
|---|---|
| (i) Very highly expressed | *rpmG* (0.729) *rplA* (0.769) *tsf* (0.770) *rplC* (0.712) *rpsG* (0.536) *tufA* (0.817) *rplL* (0.842) *fusA* (0.745) *rplQ* (0.551) *rpsT* (0.664) *lpp* (0.851) *rpmH* (0.730) *rpsA* (0.776) *rpsB* (0.774) *rplJ* (0.634) *rpsJ* (0.574) *tufB* (0.793) *ompA* (0.786) *rplK* (0.703) *rpsL* (0.661) *dnaK* (0.719) *ompC* (0.821) *rpsO* (0.688) *ompF* (0.662) *recA* (0.609) *rpmB* (0.618) *rpsU* (0.731) |
| (ii) Highly expressed | *glyS* (0.563) *rpoA* (0.447) *rpoC* (0.696) *trpS* (0.499) *rpoD* (0.587) *atpA* (0.670) *atpE* (0.585) *rpoB* (0.634) *thrS* (0.466) *tyrS* (0.508) *alaS* (0.498) *glnS* (0.556) *metG* (0.535) |
| (iii) With moderate codon usage bias | *aceE* (0.669) *asnA* (0.423) *crp* (0.490) *fumA* (0.394) *ihfA* (0.284) *malE* (0.525) *polA* (0.392) *sdhA* (0.519) *tolC* (0.391) *trxA* (0.565) *carB* (0.553) *deoC* (0.632) *dnaB* (0.394) *fimA* (0.403) *frdA* (0.556) |

|  | *ftsA* (0.323) *gdhA* (0.435) *glgC* (0.414) *nusA* (0.538) *rho* (0.544) |
| --- | --- |
| (iv) With low codon usage bias | *aroF* (0.368) *dnaQ* (0.299) *kdpB* (0.381) *lacY* (0.345) *lacZ* (0.354) *malF* (0.361) *metL* (0.398) *tonB* (0.308) *trpD* (0.353) *argI* (0.361) *atpC* (0.477) *carA* (0.401) *lysA* (0.330) *motA* (0.305) *pabA* (0.285) *phoE* (0.367) *proB* (0.361) *recF* (0.271) *tnaA* (0.492) |
| (v) Regulatory or Repressor genes | *araC* (0.252) *cytR* (0.319) *deoR* (0.331) *dnaG* (0.273) *galR* (0.331) *lacI* (0.297) *lexA* (0.379) *trpR* (0.269) |

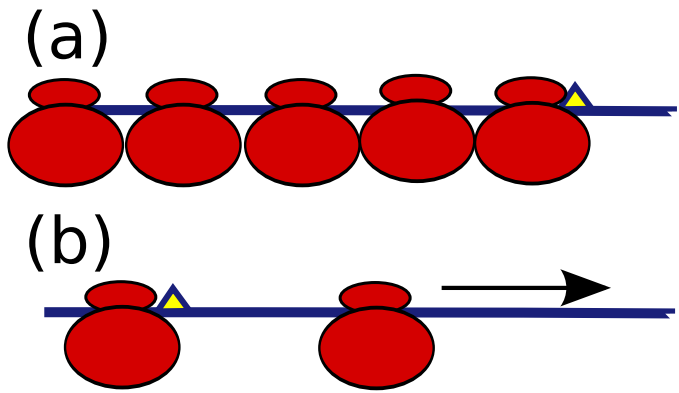

**Figure 1.** Schematic description of the effect of slow codons on ribosome traffic. (a) When there are slow codons (symbolized by a triangle obstacle) at the later part of the mRNA, many ribosomes can spend a long time in the queue. (b) When slow codons are located at the beginning, binding of ribosomes can be occluded when initiation rate is high. Ribosomes can translate quickly once they go through the slow codons, thus ribosomal efficiency is high.

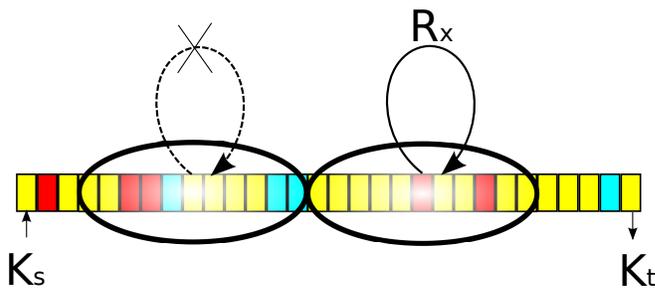

**Figure 2.** Schematic description of the ribosome traffic model. The one dimensional lattice represents mRNA, with one lattice site corresponds to one codon. A ribosome cover 11 codons and stochastically translates each codon with codon-dependent rate. Ribosomes can move forward only if the preceding site is free.

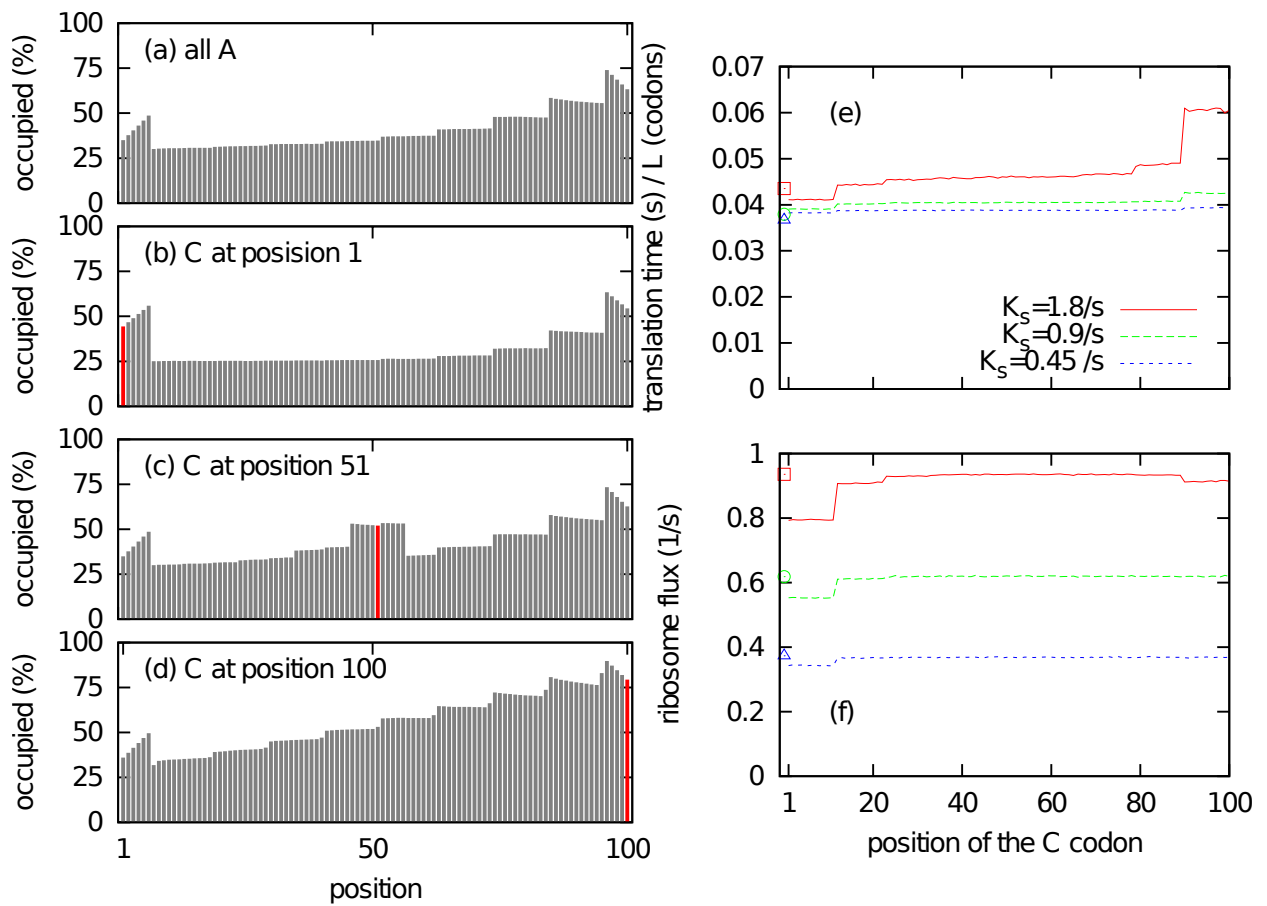

**Figure 3.** Simulation results of the artificial sequences of $L = 100$ codons, with all A-rate codons, or 99 A-rate codons and one C-rate codon. (a-d) The spatial distribution of the fraction of time for a codon to be covered by the ribosome is shown for the high on-rate $K_s = 1.8$ (1/s) for the case (a) all A-rate codons, (b) C-rate codon at position 1, (c) C-rate codon at position 51 and (d) C-rate codon at position 100. The positions of A-rate codons are shown with grey and the positions of the C-rate codons are shown with red. (e,f) The translation time per codon (the average time for a ribosome to translate the whole gene / L) (a) and the ribosome flux (b) for artificial sequences with the total length $L = 100$, with on-rate $K_s = 1.8$/s (solid lines), $K_s = 0.9$/s (dashed lines), and $K_s = 0.45$/s (dotted lines). The sequence contains 99 A-rate codons and 1 C-rate codon, and the data are plotted as the function of the position of the C-rate codon. The open symbols represent the values for the sequence with all A-rate codons, with $K_s = 1.8$/s (open squares), $K_s = 0.9$/s (open circles), and $K_s = 0.45$/s (open triangles).

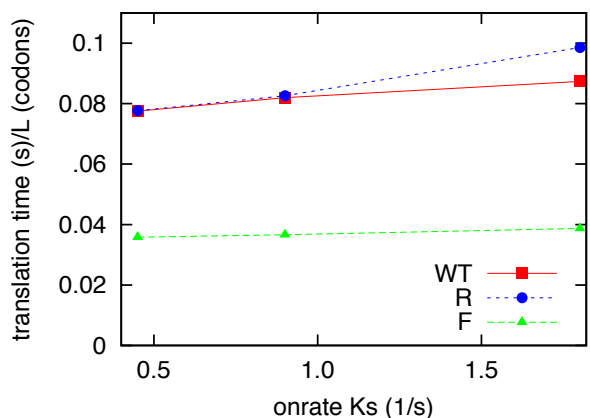

**Figure 4.** The on-rate $K_s$ dependence of the translation time (filled symbols) per codon for *lacZ* gene. WT

(squares), average from R (circles), and F (triangles) are shown.

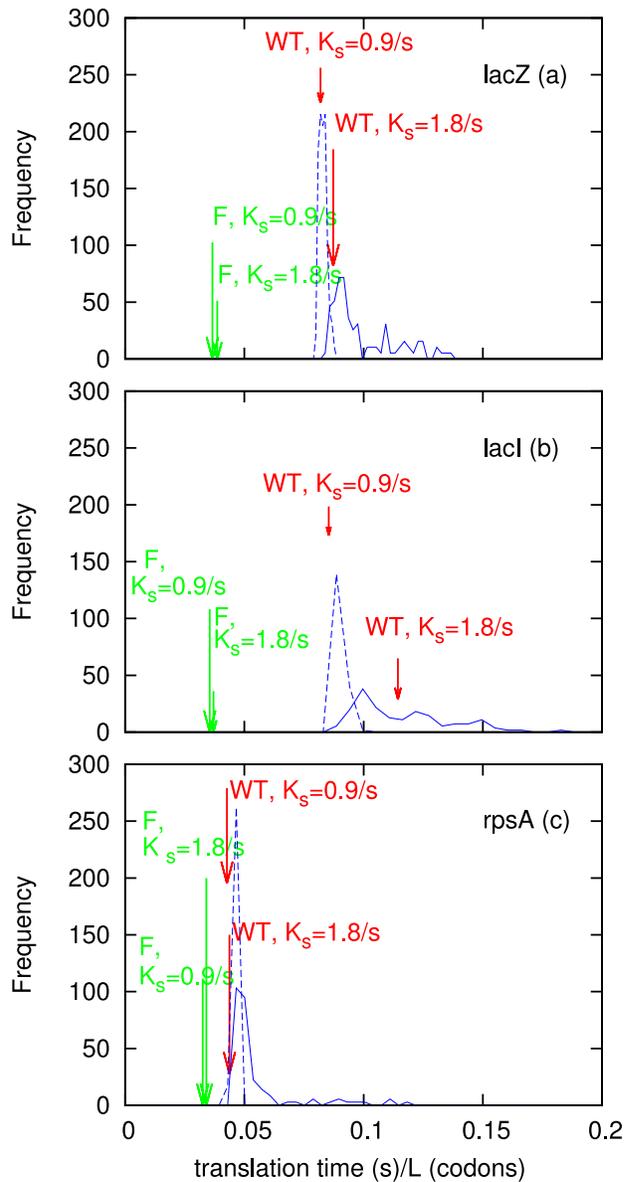

**Figure 5.** Comparison between the distributions (normalized histogram) of translation time per codon from the 100 samples of randomized sequences ("Swap") and the translation time for "WT" and "Fast" sequences for a) *lacZ* (CAI: 0.35), (b), *lacI* (CAI: 0.29) and (c) *rpsA* (CAI: 0.78). The distribution for "Swap" with the on-rate $K_s$ = 0.9/s is shown by dashed lines, the distribution for the on-rate $K_s$ = 1.8/s is shown by solid lines.

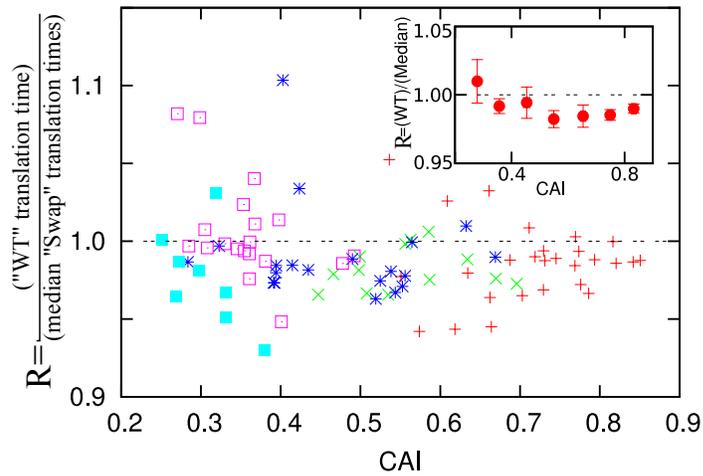

**Figure 6.** The translational efficiency ratio $R$=(WT translation time)/(Median of translation times of the 100 randomized "Swap" samples) vs. CAI for moderate on-rate $K_s$ = 0.9/s. The different categories of genes are indicated as follows: (i) Very highly expressed genes (pluses) (ii) Highly expressed genes (crosses) (iii) Genes with moderate codon usage bias (stars) (iv) Genes with low codon usage bias (open squares) (v) Regulatory / Repressor genes (filled squares). Inset shows the average of the data binned over CAI with an interval ≈ 0.1.